# Differential impact from individual versus collective misinformation tagging on the diversity of Twitter (X) information engagement and mobility

Junsol Kim[1], Zhao Wang[2], Haohan Shi[3], Hsin-Keng Ling[4] & James Evans[1,2,5] ✉

Fears about the destabilizing impact of misinformation online have motivated individuals and platforms to respond. Individuals have increasingly challenged others' online claims with fact-checks in pursuit of a healthier information ecosystem and to break down echo chambers of self-reinforcing opinion. Using Twitter (now X) data, here we show the consequences of individual misinformation tagging: tagged posters had explored novel political information and expanded topical interests immediately prior, but being tagged caused posters to retreat into information bubbles. These unintended consequences were softened by a collective verification system for misinformation moderation. In Twitter's new feature, Community Notes, misinformation tagging was peer-reviewed by other fact-checkers before revelation to the poster. With collective misinformation tagging, posters were less likely to retreat from diverse information engagement. Detailed comparison demonstrated differences in toxicity, sentiment, readability, and delay in individual versus collective misinformation tagging messages. These findings provide evidence for differential impacts from individual versus collective moderation strategies on the diversity of information engagement and mobility across the information ecosystem.

The visibility of mis- and disinformation online have attracted substantial attention around the world with demonstrations of their direct influence on major collective action in the world[1–5]. These actions range from buying and selling stocks[2] and avoidance of vaccines[3] to the attempted coup and occupation of the U.S. Capitol by rioters[4]. Legitimate fears about the destabilizing influence of false online information have inspired and put pressure on both individuals and platforms to respond. Individuals proactively correct others' claims by deploying links to fact-checking websites, such as PolitiFact and Snopes[6–10]. With the potential for amplifying misinformation through filter bubbles[11,12], social media platforms like Twitter and Facebook have come under public and political pressure to implement misinformation moderation strategies[13–15].

Individuals have become empowered to challenge others' online claims with misinformation tags (or fact-checks) in pursuit of a healthy information ecosystem and to break down ideological echo chambers[6–8]. These misinformation tags tend to target political outgroups[6,7,9], exposing tagged posters to opposing ideological perspectives. It is less clear, however, whether their misinformation tagging motivates targeted posters to explore diverse political contents afterward. Earlier research on motivated reasoning suggests that misinformation tags contradicting targeted poster's beliefs could backfire and reinforce preexisting beliefs[16,17], which could discourage people from exploring diverse information[18]. By contrast, a growing body of research argues that misinformation tagging does not backfire, but reduces engagement with misinformation and expands it with diverse information[13,14,19,20]. These mixed findings suggest that the effects of misinformation tagging could depend on the method of correcting misinformation. Individual misinformation tagging by other users often involves toxic and intolerant messages that dehumanize targeted posters[9,21], potentially hindering their willingness to explore diverse information[22].

[1]Department of Sociology, University of Chicago, Chicago, IL, USA. [2]Computational Social Science, University of Chicago, Chicago, IL, USA. [3]School of Communication, Northwestern University, Evanston, IL, USA. [4]Department of Sociology, University of Michigan, Ann Arbor, MI, USA. [5]Santa Fe Institute, Santa Fe, NM, USA. ✉e-mail: jevans@uchicago.edu

Platforms have experimented with institutionalized systems that verify the accuracy of content through collective inputs from a wider distribution of users. Notably, on Twitter's new platform, Community Notes (formerly Birdwatch), misinformation tags undergo a formal peer-review process by diverse users before being revealed to the original posters and broader Twitter user community[8,13,14]. Other platforms, including YouTube and Facebook, have recently tested or announced plans to implement features similar to Community Notes[23,24]. Rather than indiscriminately exposing users to misinformation tags, Community Notes selectively exposes misinformation tags that receive votes from heterogeneous user groups, ensuring that they are verified across a broad spectrum of perspectives[13] to activate the wisdom of crowds[25,26]. The platform also assesses the alignment of users' prior contributions with the crowd's decisions, filtering out voters who frequently oppose and backlash against valid fact-checks on misinformation. Although individual tags may be noisy and less effective, aggregating them collectively could lead to high-quality crowd judgments that align with expert fact-checks across a range of topics, from COVID-19 to politics[14,27–29]. Furthermore, the Community Notes platform has specifically instituted norms that deter toxic and intolerant misinformation tagging messages[30], potentially enhancing the efficacy of misinformation moderations and gently encouraging posters to leave their echo chambers and explore a broader world of diverse information.

In this study, we explore the impacts of individual and collective misinformation tagging on tagged posters' echo chambers. Echo chambers refer to "bounded, enclosed media spaces that have the potential to both magnify messages delivered within them and insulate them from rebuttal"[31,32], which could increase susceptibility to misinformation[11,33,34]. One indicator of echo chambers is their lack of interaction with politically diverse, cross-cutting sources of information. Prior research has measured echo chambers by selective engagement with like-minded news sources, which insulate people from opposing perspectives that could empower rebuttal[35,36]. This measure strongly correlates with other echo chamber indicators, such as intensive interactions with like-minded users (i.e., homophily)[37,38]. Literature suggests that lack of exposure to and cross-verification through opposing perspectives could erode the ability to find, evaluate, and use information effectively[11,39,40]. It could provide users with the illusion that their views are publicly supported[41,42], weakening their overall immunity against misinformation.

The other key indicator of echo chambers is their absence of content diversity resulting from limited engagement with diverse, unfamiliar topics. Emerging literature has documented the rise of socio-political endogamy, noting that both left and right increasingly develop distinct topical interests, encompassing knowledge bases, cultural tastes, and lifestyles[43–45]. For example, left-leaning individuals are more likely to engage with basic science books about physics, astronomy, and zoology, while right-leaning individuals prefer those about applied and commercial sciences like criminology, medicine, and geophysics[45]. In this way, political polarization spills over into a variety of other topics, leading to multi-dimensional segregation where opposing political groups share progressively less common ground and inhabit different realities even in topics apparently unrelated to politics[43,46]. Topical echo chambers, which magnify topics prevalent within one political group and insulate them from others, can problematize intergroup communication and interaction.

Does exposure to each type of misinformation tagging encourage or discourage posters from exploring diverse information and breaking out of echo chambers? To answer this question, we use large-scale digital traces from the platform formerly known as Twitter (X as of July, 2023) to identify posters exposed to each approach of misinformation tagging. First, we identify posters targeted by individual misinformation tags. These posters' tweets received other individuals' voluntary replies, citing fact-checking articles from PolitiFact, one of the largest and most studied professional fact-checking organizations in the United States[7,10]. Second, we examine posters targeted by collective misinformation tags. These posters' tweets received notes that contain collectively verified fact-checks through Twitter's Community Notes platform. Figure 1a visualizes the mechanism of each type of misinformation tagging, which represent the most prevalent misinformation moderation strategies on Twitter[6–10,13–15]. Supplementary Fig. 1 presents an example of individual and collective tags that correct topically identical, COVID-19 misinformation.

Using 712,948 tweets that cite news sources—including posts, retweets, and quotes—posted by 7733 users before and after they were targeted by misinformation tags, we estimate the effects of these tags on the posters' echo chambers. Specifically, we measure echo chambers using political and content diversity in their posting and sharing behavior (see Fig. 1b). Political diversity measures whether a poster's tweet cites a source with opposing political stance (e.g., a right-leaning poster references left-leaning articles)[5,47]. Content diversity measures whether a tweet discusses novel topics unfamiliar in the poster's historical tweets. We apply a transformer-based sentence embedding model (SentenceBERT) to extract a high-dimensional, semantic vector representation for each tweet, and aggregate the vectors of each author's historical tweets to produce an average semantic vector for each poster. We then measure the distance between a particular tweet and the poster to assess the degree to which this tweet expands the poster's content diversity. As our data focus on tweets citing news sources, we assume that the increase of content diversity indicates the exploration of novel political news topics. For example, consider a user who regularly consumes and shares news about COVID-19 but begins to discuss U.S. tax and labor issues as well. This shift indicates an increase in the user's content diversity, as detailed in Supplementary Table 1. We consider both political and content diversity because they represent different dimensions that could reinforce one another in limiting exposure to information and exacerbating echo chambers on social media[17,43,48].

## Results

We aim to investigate the effects of individual and collective misinformation tagging on political and content diversity using large-scale Twitter data. In our observational data, treatments (i.e., exposure to misinformation tagging), however, are not randomly assigned to misinformation posters, which pose challenges for identifying the causal effects of misinformation tagging. To address these concerns, we apply interrupted time series (ITS) and delayed feedback (DF) analysis, which help eliminate non-causal explanations under certain assumptions.

### Interrupted time series (ITS) analysis

Interrupted Time Series (ITS) analysis investigates whether the trend in political and content diversity shifts after misinformation tagging. ITS assumes that without the intervention of misinformation tagging, the pre-treatment trend (i.e., before misinformation tagging) would persist, and the immediate change in trend after misinformation tagging is attributed to effects from tagging. We control for user-level fixed effects to correct for time-invariant user characteristics.

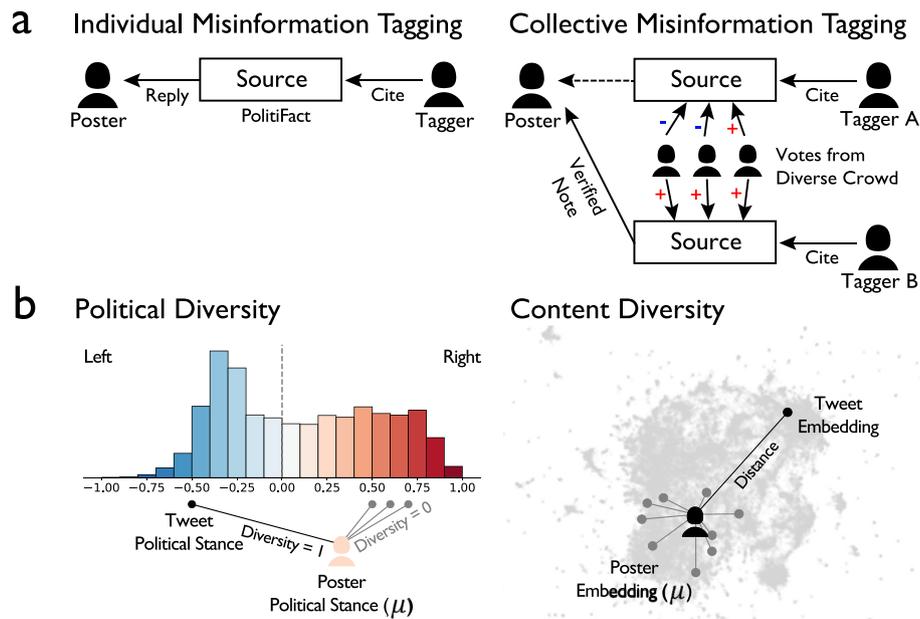

**Fig. 1 | Misinformation Tagging and Outcomes Measurement. a** Individual misinformation tagging in which individuals cite PolitiFact fact-checking articles. Collective misinformation tagging through the Community Notes platform, which selectively exposes verified misinformation tags that receive diverse votes as helpful. **b** Operationalization of tweet political and content diversity. Political diversity captures whether a poster cites a source with opposing political stance (binary 0/1), assessed from the aggregate stances of referenced sources. Content diversity captures whether a post discusses topics unfamiliar to the author's historical tweets (continuous), assessed with the distance between the poster's average tweet and a particular tweet within a contextual embedding (sentenceBERT pre-trained on Twitter)[74].

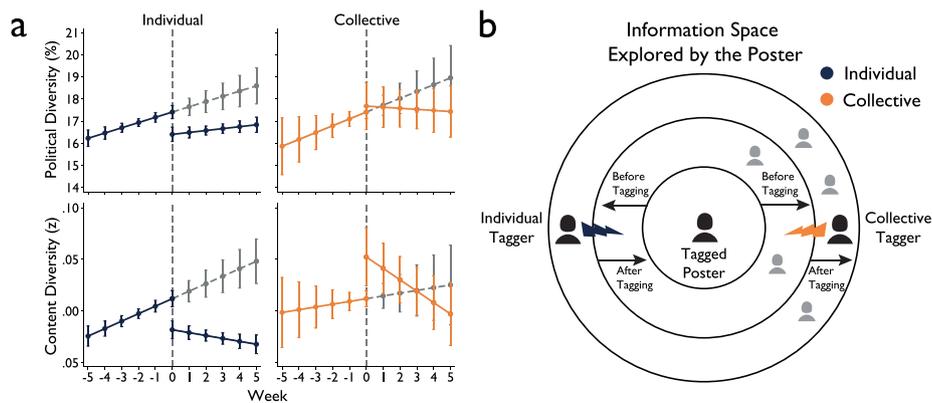

**Fig. 2 | Political and content diversity change with the intervention of individual and collective misinformation tagging. a** Results from Interrupted Time Series (ITS) analysis. The *x*-axis denotes the timeline of tweets posted before and after tagging, with negative values representing the number of weeks before posting tagged tweets and positive values representing the number of weeks after. The *y*-axis represents political and content diversity, with dots indicating the diversity score for each corresponding week as estimated by the ITS analysis, and error bars showing 95% confidence intervals. Solid lines connect the dots revealing trends of political and content diversity before and after tagging, with gray dotted lines tracing the counterfactual trend if fact-checks had not occurred. The sample size is 424,969 tweets. There is no control group in this analysis; however, a comparative interrupted time series analysis with a control group can be found in Supplementary Method 3. **b** Illustration of political and content diversity dynamics before and after tagging. Before individual and collective tagging, posters exhibit increased political and content diversity, which increases the likelihood of encountering a fact-checker. After individual tagging, posters retreat into information bubbles; after collective tagging, they venture further beyond them.

Figure 2a and Table 1 report results from our ITS analysis (Political Diversity: $R^2 = 0.173$, Content Diversity: $R^2 = 0.243$). Posters manifest an increasing tendency to explore novel political information before being fact-checked by misinformation tags. Specifically, before individual and collective misinformation tagging, posters increase the political diversity ($\beta = 0.237$, 95% CI = [0.125, 0.349], $t(418115) = 4.14$, $p < 0.001$) and content diversity ($\beta = 0.007$, 95% CI = [0.004, 0.010], $t(418115) = 4.79$, $p < 0.001$) of their information engagement over time.

Having their posts criticized by individual misinformation tags, however, causes posters to retreat within an information bubble. Immediately after tagging, posters significantly decrease the political diversity ($\beta = -1.009$, 95% CI = [−1.447, −0.571], $t(418115) = -4.52$,

Table 1 | Interrupted Time Series (ITS) Model Results for Political and Content Diversity

| Outcome | Political diversity (%) | | | Content Diversity (z) | | |
|---|---|---|---|---|---|---|
| Type of misinformation tagging | Individual | Collective | Difference (Collective - Individual) | Individual | Collective | Difference (Collective - Individual) |
| Slope before posting the tweet | 0.237*** [0.125, 0.349] t = 4.14, p < 0.001 | 0.309* [0.041, 0.578] t = 2.26, p = 0.024 | 0.072 [−0.219, 0.363] t = 0.480, p = 0.628 | 0.007*** [0.004, 0.010] t = 4.79, p < 0.001 | 0.003 [−0.004, 0.010] t = .74, p = 0.461 | −0.005 [−0.012, 0.003] t = −1.17, p = 0.243 |
| Immediate intercept change after tagging | −1.009*** [−1.447, −0.571] t = −4.52, p < 0.001 | 0.270 [−0.824, 1.363] t = 0.48, p = 0.629 | 1.279* [0.101, 2.457] t = 2.13, p = 0.033 | −0.030*** [−0.042, −0.019] t = −5.10, p < 0.001 | 0.040** [0.012, 0.069] t = 2.74, p = 0.006 | 0.070*** [0.039, 0.102] t = 4.44, p < 0.001 |
| Slope after tagging | 0.087 [−0.020, 0.194] t = 1.60, p = 0.110 | −0.049 [−0.334, 0.235] t = −0.34, p = 0.734 | −0.136 [−0.440, 0.167] t = −0.88, p = 0.379 | −0.003 [−0.0006, 0.000] t = −1.95, p = 0.051 | −0.011** [−0.019, −0.004] t = −2.88, p = 0.004 | −0.008* [−0.016, 0.000] t = −2.01, p = 0.044 |
| Slope change (After - Before) | −0.150 [−0.306, 0.006] t = −1.89, p = 0.059 | −0.358 [−0.749, 0.033] t = −1.80, p = 0.072 | −0.208 [−0.629, 0.213] t = −0.97, p = 0.332 | −0.010*** [−0.014, −0.006] t = −4.79, p < 0.001 | −0.014** [−0.024, −0.003] t = −2.60, p = 0.009 | −0.004 [−0.015, 0.007] t = −0.64, p = 0.520 |
| $R^2$ | 0.173 | | | 0.243 | | |
| Adjusted $R^2$ | 0.159 | | | 0.230 | | |
| Observations | 424,969 | | | | | |
| df | 418,115 | | | | | |

Notes: ***p < 0.001 **p < 0.01 *p < 0.05. We multiply political diversity by 100 to interpret the estimates as absolute percentage point changes. We normalize content diversity to z-scores (the number of standard deviations from the mean). All regressions control for user fixed effects and the number of tweets per day. The statistical significance of regression coefficients is tested using two-sided t-tests. Confidence intervals (95%) are provided in brackets, along with the corresponding t-statistics, degrees of freedom, and exact p-values. More details can be found in Methods: Interrupted Time Series (ITS) Analysis.

$p < 0.001$) and content diversity ($\beta = -0.030$, 95% CI = [−0.042, −0.019], $t(418115) = -5.10$, $p < 0.001$) of their posts. After tagging, the slope becomes nearly flat, indicating that posters' future posts continue to collapse in both political diversity ($\beta = 0.087$, 95% CI = [−0.020, 0.194], $t(418115) = 1.60$, $p = 0.110$) and content diversity ($\beta = -0.003$, 95% CI = [−0.006, 0.000], $t(418115) = -1.95$, $p = 0.51$).

By contrast, there is no statistically significant evidence that collective misinformation tagging causes individuals to retreat within their prior information bubble. The data even reveals a slight, although not significant, increase in political diversity ($\beta = 0.270$, 95% CI = [−0.824, 1.363], $t(418115) = 0.48$, $p = 0.629$) and a significant increase in content diversity ($\beta = 0.040$, 95% CI = [0.012, 0.069], $t(418115) = 2.74$, $p = 0.006$) immediately after tagging. Nevertheless, collective misinformation tagging has only a temporary effect on individual posters. Especially, the slope for content diversity changes significantly after tagging ($\beta = -0.014$, 95% CI = [−0.024, −0.003], $t(418115) = -2.60$, $p = 0.009$), eventually converging to levels experienced before the initial misinformation tags occur. Despite the steepness of the slope following collective tagging, our analysis indicates that content diversity does not significantly drop below the pre-tagged period (see Supplementary Method 1).

We find that the gap between the effects of individual and collective misinformation tagging is significant, particularly regarding the immediate intercept change in political diversity ($\beta_{Individual} = -1.009$, $\beta_{Collective} = 0.270$, $\beta_{Collective} - \beta_{Individual} = 1.279$, 95% CI = [0.101, 2.457], $t(418115) = 2.13$, $p = 0.033$) and in content diversity ($\beta_{Individual} = -0.030$, $\beta_{Collective} = 0.040$, $\beta_{Collective} - \beta_{Individual} = 0.070$, 95% CI = [0.039, 0.102], $t(418115) = 4.44$, $p < 0.001$).

Additional analyses reveal the effects of misinformation tagging on the proximity between posters and misinformation taggers. This suggests that Twitter navigation likely makes posters more visible to fact-checkers as they venture into foreign territory (see Fig. 2b). Exposure to fact-checks causes them to retreat back into their information bubbles, distancing them from the foreign stances that fact-checked them (see Supplementary Method 2).

Because time-variant confounders (e.g., viral news, platform algorithm changes, or significant external events) can affect ITS outcomes, we conduct additional analyses to control for these factors. First, we control for major events during the study period through sensitivity analyses. Second, we apply comparative interrupted time series (CITS) analyses. These additional analyses support our initial findings (see Supplementary Method 3). Additionally, to address autocorrelated posting behaviors among social media users, we include autoregressive terms in the ITS models, further enhancing the robustness of our findings (see Supplementary Method 4).

To better understand what happens when posters retreat to their information bubbles, we conduct a series of descriptive analyses (see Supplementary Table 2). When posters reduce their political and content diversity, the number of tweets (comprising posts, retweets, and quotes) posted per day significantly increases, indicating that users are more active within their information bubbles. Specifically, the number of tweets per day is negatively correlated with political diversity ($r = -0.107$, $t(712946) = -90.87$, 95% CI = [−0.109, −0.105], $p < 0.001$) and content diversity ($r = -0.052$, $t(712946) = -43.97$, 95% CI = [−0.054, −0.050], $p < 0.001$). Similarly, we find that the type of posting is different; the proportion of retweets (i.e., tweets simply sharing other users' tweets) out of the entire tweets per day is negatively correlated with political diversity ($r = -0.046$, $t(712946) = -38.88$, 95% CI = [−0.048, −0.044], $p < 0.001$) but positively correlated with content diversity ($r = 0.012$, $t(712946) = 10.13$, 95% CI = [0.010, 0.014], $p < 0.001$). This indicates that users actively post tweets rather than passively retweet other users' tweets when they exhibit low political diversity. To demonstrate the significant effects of misinformation tagging on political and content diversity, irrespective of these factors, we have adjusted for the number of tweets posted per day. We have also controlled for the proportion of retweets per day, which did not meaningfully change our results (see Supplementary Table 3).

**Delayed feedback (DF) analysis**
We employ delayed feedback (DF) analysis to further strengthen our causal inference[49]. In our DF analysis, we estimate baseline changes (i.e., changes in outcomes that occur without tags) to answer the question: "Are shifts in political and content diversity attributable to tagging, or do similar changes occur even without tagging?" Pairs of tweets containing similar misinformation, targeted by misinformation tagging at different times, are matched to construct a control group, consisting of posters whose problematic tweets have not yet been tagged due to delayed feedback, and a treatment group of posters who have. For instance, Supplementary Fig. 2 presents an illustrative example involving a pair of matched tweets and tags.

In Fig. 3a, post-treatment ($t_1$) represents the time window when treatment tweets are tagged but control tweets are not, and pre-treatment ($t_0$) represents the time window with equal duration $t_1$ when both treatment and control tweets are untagged. Changes in the outcomes between $t_0$ and $t_1$ in the control group reflect baseline changes, which indicate changes without tags. Changes between $t_0$ and $t_1$ in the treatment group reflect treated changes, which indicate changes with tags. We compare the difference in pre-post change between control and treatment groups (i.e., baseline vs. treated changes) to identify the effects of misinformation tagging on political and content diversity. DF analysis assumes that, in the absence of treatment, both control and treatment groups would exhibit parallel trends. We control for user-level fixed effects to control for time-invariant, user-specific characteristics.

Figure 3b and Table 2 present results from the DF analysis (Political Diversity: $R^2 = 0.274$, Content Diversity: $R^2 = 0.358$). Our DF analysis demonstrates that changes are indeed due to tagging, showing that treated changes are significant above and beyond baseline changes. Consistent with the ITS findings, DF analysis indicates that individual misinformation tags lead to a significant decrease in political diversity ($\beta = -5.886$, 95% CI = [−9.633, −2.138], $t(8182) = -3.08$, $p = 0.002$). Nevertheless, individual misinformation tagging does not significantly affect content diversity ($\beta = 0.018$, 95% CI = [0.145, 0.403], $t(8182) = 4.17$, $p = 0.652$). Although ITS analyses show that content diversity decreases after tagging, DF analyses indicate no statistically significant evidence that content diversity decreases beyond baseline changes observed without tags. Collective misinformation tags, by contrast, do not produce a significant decrease in political diversity ($\beta = 1.219$, 95% CI = [−4.777, 7.215], $t(8182) = 0.40$, $p = 0.690$) and even increase content diversity following tagging ($\beta = 0.274$, 95% CI = [0.145, 0.403], $t(8182) = 4.17$, $p < 0.001$). The gap between the effects of individual and collective tagging is significant for both political diversity ($\beta = 7.105$, 95% CI = [0.069, 14.140], $t(8182) = 1.98$, $p = 0.048$) and content diversity ($\beta = 0.256$, 95% CI = [0.105, 0.407], $t(8182) = 3.32$, $p = 0.001$).

### Linguistic characteristics of misinformation tags

Individual and collective misinformation tagging messages manifest different linguistic characteristics. As shown in Fig. 4 and Supplementary

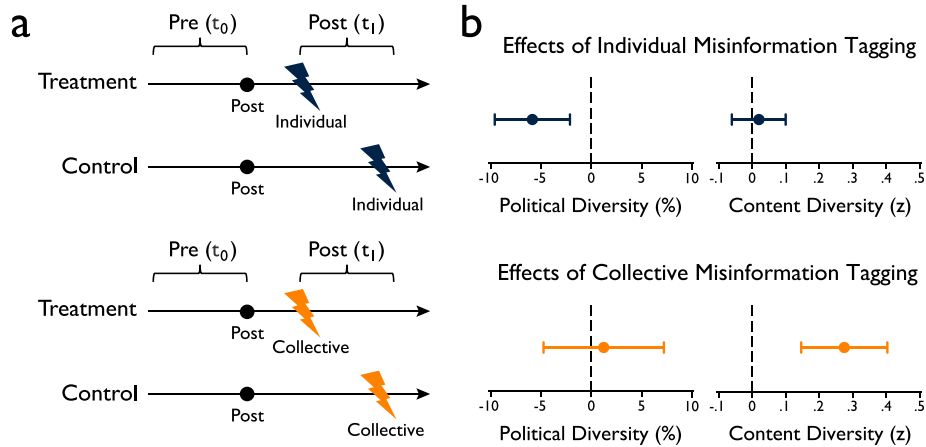

**Fig. 3 | Delayed feedback (DF) analysis. a** Pre- and post-treatment periods. Post-treatment ($t_1$) represents the time window when treated tweets are tagged but control tweets are not. Pre-treatment ($t_0$) represents the time window with equal duration $t_1$ when both treatment and control tweets remain untagged. **b** The effects of individual and collective misinformation tagging on political and content diversity are estimated by the difference in pre-post changes in outcomes between the treatment and control groups. Dots represent the difference in pre-post changes for each outcome between the treatment and control groups, with error bars indicating 95% confidence intervals. Pairs of tweets containing similar misinformation, targeted by misinformation tagging at different times, are matched to construct the control group. The control group consists of users whose problematic tweets had not yet been tagged due to delayed feedback, while the treatment group consists of users whose problematic tweets had already been tagged (see "Methods": Delayed Feedback (DF) Analysis for details). The sample size is 8901 tweets.

**Table 2 | Delayed feedback (DF) model results for political and content diversity**

| Outcome | Political diversity | | | Content Diversity | | |
|---|---|---|---|---|---|---|
| Type of misinformation tagging | Individual | Collective | Difference (Collective - Individual) | Individual | Collective | Difference (Collective - Individual) |
| Difference in Pre-Post Change (Treatment - Control) | −5.886** [−9.633, −2.138] $t = -3.08$, $p = 0.002$ | 1.219 [−4.777, 7.215] $t = 0.40$, $p = 0.690$ | 7.105* [0.069, 14.140] $t = 1.98$, $p = 0.048$ | 0.018 [−0.062, 0.099] $t = 0.45$, $p = 0.652$ | 0.274*** [0.145, 0.403] $t = 4.17$, $p < 0.001$ | 0.256** [0.105, 0.407] $t = 3.32$, $p = 0.001$ |
| $R^2$ | 0.274 | | | 0.358 | | |
| Adjusted $R^2$ | 0.211 | | | 0.301 | | |
| Observations | 8901 | | | | | |
| df | 8182 | | | | | |

Notes: ***$p < 0.001$ **$p < 0.01$ *$p < 0.05$. Each cell presents the difference in pre-post change (treatment group - control group) in each outcome. We multiply political diversity by 100 to interpret the estimates as absolute percentage point changes. We normalize content diversity to z-scores (the number of standard deviations from the mean). All regressions control for user fixed effects and the number of tweets per day. The statistical significance of regression coefficients is tested using two-sided t-tests. Confidence intervals (95%) are provided in brackets, along with the corresponding t-statistics, degrees of freedom, and exact p-values. More details can be found in Methods: Delayed Feedback (DF) Analysis.

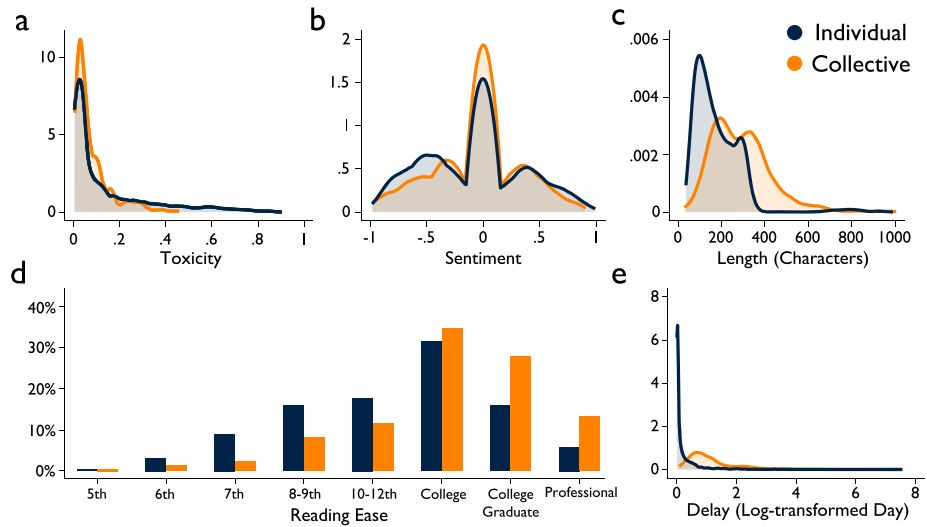

**Fig. 4 | Linguistic characteristics of fact-checking messages. a** univariate kernel density function for toxicity. **b** univariate kernel density function for sentiment. **c** univariate kernel density function for length (characters). **d** histogram for reading ease (**e**) univariate kernel density function for delay (log-transformed days). The purple line (or bar) represents the distribution within individual misinformation tags, while the yellow line (or bar) represents the distribution within collective tags.

Table 4, we find that individual misinformation tags exhibit twice the toxic content (Mean$_{Individual}$ = 0.139, Mean$_{Collective}$ = 0.076, Mean$_{Collective}$-Mean$_{Individual}$ = −0.063, $t(7496)$ = 9.86, $p < 0.001$, Cohen's $d$ = −0.228) and convey more negative sentiment compared to collective misinformation tags (Mean$_{Individual}$ = −0.082, Mean$_{Collective}$ = −0.050, Mean$_{Collective}$-Mean$_{Individual}$ = 0.032, $t(7731)$ = 2.14, $p = 0.033$, Cohen's $d$ = 0.049). Collective tags express slightly higher positive sentiment and produce messages with more neutral sentiment than individual tags. Furthermore, individual tag messages are much shorter (Mean$_{Individual}$=179.31, Mean$_{Collective}$ = 288.87, Mean$_{Collective}$-Mean$_{Individual}$ = 109.56, $t(7731)$ = 26.95, $p < 0.001$, Cohen's $d$ = 0.613) and more readable ($\chi^2(7)$ = 155.32, $p < 0.001$, Cramer's $V$ = 0.155) than collective tags. While 53.53% of individual tags necessitate a college-level reading comprehension or higher, 75.77% of collective tags demand this level. Moreover, the delay between posting misinformation and receiving fact-checks is shorter for individual than collective tagging (Mean$_{Individual}$=3.037, Mean$_{Collective}$ = 6.322, Mean$_{Collective}$-Mean$_{Individual}$ = 3.285, $t(7731)$ = 2.13, $p = 0.033$, Cohen's $d$ = 0.048). These findings demonstrate that individual tags convey their messages quickly through messages that are succinct, straightforward, emotive, and sometimes toxic. In contrast, collective tags are more slowly communicated through lengthy, complex messages, devoid of emotional undertone or toxicity.

Based on linguistic differences between individual and collective tags, we question whether gaps in the effects of individual versus collective tags persist, even when the linguistic characteristics of these tags are similar. First, we control for toxicity by excluding tags with a toxicity level higher than 0.4 and retrain only non-toxic tags. Second, we control for sentiment by removing tags with either positive (> 0.2) or negative (< −0.2) sentiments, keeping only neutral tags. Third, we control for length by excluding tags longer than 400 characters and retaining short tags. Fourth, we control for readability by excluding tags that require college-level or higher readability and selecting tags that are relatively easy to read. Fifth, we control for delay by omitting any tags associated with delays longer than 48 hours (log-transformed delay > 1.10) and focusing on quick tags.

We find that the gap between individual and collective tagging remains statistically significant, except when controlling for length. As shown in Supplementary Table 5, the gap in political diversity is not statistically significant after controlling for length ($\beta = 1.071$, 95% CI = [−0.231, 2.373], $t(399236) = 1.61$, $p = 0.107$). Nevertheless, controlling for length only accounts for 16.26% of the gap between individual and collective tagging in political diversity. This indicates that linguistic characteristics explain a modest but nontrivial portion of the differential impacts between individual and collective tagging. Nevertheless, these measured qualities do not account for the vast majority of the difference.

### Control analyses

In this section, we identify systematic differences in misinformation that receive individual versus collective tagging, as well as differences in the posters corrected by each type. Even after controlling for these differences in additional interrupted time series (ITS) analyses, individual and collective tagging have significantly different effects in the directions identified by our unconstrained analysis.

First, we observe that individual taggers focus more on political topics, while collective taggers correct a more diverse range of topics (see Supplementary Table 6). As shown in Supplementary Table 7, the nine most frequent topics in our dataset include political topics known to trigger divisive, polarized reactions in US politics (see "Methods": Topic modeling). These topics account for 84.06% of the corrections made through individual tagging but only 59.49% of the corrections made through collective tagging.

Therefore, we control for topics of the corrected misinformation, finding that the gaps between individual and collective tags are significant and even slightly larger when they correct identical topics of misinformation. Specifically, we employ propensity score weighting (PSW) method (see Supplementary Method 5). The results demonstrate that even when individual and collective tagging correct topically identical messages, the gap between individual and collective tagging is significant, both in the immediate change of political diversity ($\beta = 2.380$, 95% CI = [0.200, 4.560], $t(296544) = 2.14$, $p = 0.032$) and content diversity ($\beta = 0.048$, 95% CI = [0.003, 0.092], $t(296544) = 2.11$, $p = 0.035$). We note that collective tagging is less likely to correct political topics than individual tagging but is more effective in causing original posters to explore diverse content when successfully deployed on political topics. Refer to Supplementary Table 8 for details.

Second, we find that the proportion of right-leaning users corrected by individual tagging is 53.17% while right-leaning users corrected by collective tagging is 44.14%. We also analyze the distribution of political stance among taggers (i.e., those who write individual tags) and voters (i.e., those who vote on the exposure of collective tags) (see Supplementary Method 6 and Supplementary Table 9). We compare the effects of individual and collective tagging in the common scenario where right-leaning posters are corrected by left-leaning ones. Specifically, we focus on cases where right-leaning posters are corrected either by individual tags from left-leaning taggers or by collective tags approved by a majority of left-leaning voters (i.e., Community notes approved by voters, where at least 50% of those with identifiable political stances are left-leaning). In this analysis, the difference between the effects of individual and collective tagging is still significant, both in the immediate change of political diversity ($\beta = 1.780$, 95% CI = [0.118, 3.441], $t(238081) = 2.10$, $p = 0.036$) and content diversity ($\beta = 0.076$, 95% CI = [0.033, 0.119], $t(238081) = 3.46$, $p = 0.001$). Refer to Supplementary Table 10 for details.

Third, we find that popular users are more likely to receive collective tags than individual tags, which is consistent with prior literature (see Supplementary Fig. 3)[8]. To examine the differences between individual and collective tags when focusing on less popular, everyday users, we exclude those whose number of followers exceeds 2967, the average number of followers among users corrected by individual tags. We find the results are consistent overall (see Supplementary Table 11), but suggest that collective tagging of low popularity posters is slightly more effective, relative to individual tagging, than with high popularity users. In particular, the difference between individual and collective tagging is significant, both in the immediate change of political diversity ($\beta = 3.612$, 95% CI = [0.824, 6.399], $t(235632) = 2.54$, $p = 0.011$) and content diversity ($\beta = 0.081$, 95% CI = [0.000, 0.162], $t(238081) = 1.97$, $p = 0.049$). This may indicate the inoculation of popular users to critique, an increased sensitivity among unpopular users to collective nudges[50], or both.

**Robustness checks**

We verify our findings with a battery of robustness checks. First, we seek to avert concerns over the presence of bots on Twitter by reanalyzing our data excluding identified bot accounts[2,5]. Second, we reanalyze the relationship controlling for potentially insincere informational activities, such as citing sources of low credibility and intentionally spreading fake news. Third, we attempt to avoid situations in which posters simply criticize distant information without honest consideration by filtering out posts with negative sentiment. Fourth, we identify all tweets within the sample that mention keywords related to receiving community notes broadly and remove them, as they could confound our measure of content diversity. To address concern regarding the effect of replying directly to individual taggers, which could confound the measure of political diversity, we also identify and remove all tweets that reply directly to individual taggers. Fifth, to strictly identify individual tags (i.e., PolitiFact links) that correct the original posters, we prompt ChatGPT to annotate whether the links are used to correct the original poster rather than support them. Then, we limit the sample to links that correct the original posters. Sixth, considering the low visibility of individual tags in Twitter's message-reply interface[6,8], we restrict the sample to original posters who replied to (and thereby read) the individual tags and remove non-responders. These alterations do not meaningfully impact our reported outcomes (see "Method": Robustness Checks).

**Discussion**

This study provides empirical evidence regarding the impact of individual and collective misinformation tagging on echo chambers. Before misinformation tagging, posters show an increased curiosity in diverse political and topical content. This challenges the conception that misinformation is generated and corrected when people retreat into echo chambers[11,33]. On the contrary, posters become fact-checked when they venture outside those bubbles. Why is exploration followed by misinformation tagging? First, posters could misinterpret unfamiliar and diverse information from a lack of information literacy[51], increasing the chance of posting the misinformation being tagged. Second, news feed algorithms may increase the probability that posters' tweets become visible to people from political outgroups, who are highly motivated to fact-check foreign posters[6,7,14]. Our analysis shows that posters increase the closeness to misinformation taggers before fact-checks, which could increase the chance of appearing in fact-checkers' news feeds.

Individual misinformation tagging discourages posters from exploring diverse information. Posters tagged by individuals manifest an immediate drop in political diversity, as evidenced by both interrupted time series (ITS) and delayed feedback (DF) analyses. Content diversity also decreases in ITS analyses, although DF analyses do not reveal a significant drop. This suggests that while content diversity decreases after tagging, it does not fall below the baseline change expected without tags. These unintended consequences are mitigated by collective misinformation tagging. Unlike individual tagging, there is no statistically significant evidence that collective tagging diminishes political and content diversity in both ITS and DF analsyzes; moreover, it results in a short-term rise in content diversity.

Our analyses show that individual tagging involves short, toxic, and emotion-driven messages. Collective tagging, on the other hand, involves longer, less toxic, emotionally neutral, and deliberative messages revealed to posters longer after their offending posts. These results suggest the trade-off between the effectiveness of established systems for promoting openness and mobility across the information ecosystem, but the efficiency of individuals in cleaning it. Low visibility of individual misinformation tagging in Twitter's message-reply interface[6,8] may motivate taggers to use short and potentially toxic messages. Community Notes responded by implementing a more visible interface for collaborative tagging, which reduces the tendency to terseness, facilitating long and deliberate discussion. Also, norms and values underlying participation in Community Notes could prevent taggers from disseminating succinct yet inflammatory messages viewed as unhelpful and instead source diverse perspectives[13].

What mechanisms drive differences in the effects of individual and collective misinformation tagging on echo chambers? We find that linguistic characteristics, such as toxicity, sentiments, and length only partially explain differential impacts between individual and collective tagging. This implies that differences in quality other than linguistic characteristics also exert a direct influence. Literature on the wisdom of crowds suggests that while individual tags are susceptible to biases and noise, aggregating tags collectively could correct individual bias, increasing the quality of nonexpert fact-checks[28,52,53]. For example, compared to individual tags, collective tags are more closely aligned with professional fact-checks from experts on a variety of topics, ranging from COVID-19 to politics[14,27–29]. Even though we focus on individual tags that cite professional fact-checks (i.e., PolitiFact), it is possible that interpretations within individual tags might be less effective when not cross-validated like collective tags. For example, individual tags might fail to convey the key points of PolitiFact articles or clearly articulate the relevance of these articles to the original post. Additionally, when multiple fact-checkers co-validate collective tags, these decisions may be perceived as more legitimate and less susceptible to biases, encouraging the original posters to seek out more diverse and cross-validating information[28].

Overall, our findings suggest that misinformation is posted and fact-checked when original posters who were accustomed to like-minded sources associated with low credibility (see Supplementary Table 2) suddenly increase their political and content diversity. In the short term, some might believe that pushing them back into their echo chambers with individual tags seems like an effective way to curb misinformation. Nevertheless, over the long term, this approach could expand the cluster of users immersed in misinformation, depriving them of opportunities to educate themselves with opposing perspectives. The ethical and normative aspects of our research remain open questions, but we suggest that collective tagging encouraging exploration might be better for the long-term health of the information ecosystem.

Our analyses have several notable limitations. First, our method for assessing posters' political stances is indirect, through their posting behavior[5]. This approach has been successfully applied to predict political party affiliation and self-described ideology in previous literature[53], but using a direct measure of political ideology or affiliation with social media and survey data would strengthen our assessments. Second, our quasi-experimental methodologies (ITS and DF) depend on assumptions for causal inference. We employ topic modeling and matching to enhance tweet comparability within treatment and control groups, but acknowledge that unobserved time-variant confounders may influence posters' responses. Third, although we have employed a popular bot detection algorithm, recent studies have suggested that algorithmic removal of bots is challenging and may introduce additional bias[54]. Therefore, we report the full results with and without the algorithmic removal of bots, demonstrating that our results are consistent. To thoroughly remove bots, future research could match social media data with survey or administrative data (e.g., voter records) to ensure the authenticity of participants[55]. Fourth, Twitter (X as of July, 2023) closed access to the Academic Research API, which had been freely available to eligible researchers until May 2023. This could limit other researchers' ability to reproduce our findings with recent data after May 2023[56]. Collective tagging systems are increasingly being deployed across social media platforms, such as Twitter's Community Notes and similar features currently being tested on platforms like Facebook and YouTube[23,24]. Future research should examine whether our findings are reproducible across different platforms, time periods, and cultural contexts. Fifth, we employ the topic modeling and propensity score weighting (PSW) method to control for semantic differences between tweets tagged by individual and collective tagging (refer to Supplementary Method 5). Nevertheless, PSW might fail to address the confounding effects of unobserved semantic differences beyond topics. Despite these limitations, our study uncovers a significant and substantial relationship between fact-checks and reduced information diversity. We also demonstrate the power of designed institutions, like collective fact-checking on Twitter, to moderate the negative, narrowing effects of fact-checking on information exploration.

## Methods

### Data

Our study complied with the terms of all data sources used in the study (including but not limited to Twitter/X). Using the Twitter API v2.0 with academic research access, we collected Twitter data to explore the effects of individual and collective misinformation tagging. First, we identified 9,372 users targeted by individual misinformation tagging from 2021/10/1 to 2022/3/25. We selected users whose tweets received fact-checking replies that contain URLs to fact-checking articles from "politifact.com." Second, we identified 1,465 users targeted by collective tagging from 2022/12/19 to 2023/3/31, when Community Notes were made public to Twitter users globally[57]. In Community Notes, users can flag any tweets as misinformation with notes, and other members vote for the helpfulness of the notes. (Users also have the option to flag tweets they believe are free from misinformation; however, these instances have been excluded from our analysis.) Collectively verified notes that received the above-threshold helpfulness votes from a diverse set of users are then made public to the original user (who posted the misinformation) and the broad Twitter audience[13]. In our work, we only considered notes with above-threshold helpfulness votes. Note that the platform also assesses the alignment of users' prior contributions with the crowd's decisions, filtering out voters who frequently oppose and backlash against valid fact-checks on misinformation (see Supplementary Method 7).

Due to the rate limit of Twitter API, we only collected data from regular Twitter users, excluding organizations' and celebrities' accounts with 50,000 or more followers. Additionally, to focus on individual users, rather than organizational accounts (e.g., CNN, Fox News, etc), we removed 1,659 users identified as organization accounts by the M3Inference library[58,59]. We further removed 1445 users who were fact-checked more than once within the period of data collection to avoid the potential for them to become desensitized for repeated fact-checks. After filtering the data, our final dataset included 7733 users, where 6760 users were targeted by individual misinformation tagging and 973 users were targeted by collective misinformation tagging. We found that individual tagging is more frequent than collective tagging in our dataset due to the cross-validation process required to expose collective tags. This leads to an imbalance in group size between users corrected by individual and collective tags. Nevertheless, our statistical models (interrupted time series and delayed feedback models) do not assume equal group size for comparison between the effects of individual and collective tagging. Also, we found that 16.33% of tweets that received individual tags and 15.60% of tweets that received collective tags were removed by Twitter or by the original poster. The probability of removal is similar between individual and collective tags (Difference = 0.73%, $z = 0.629$, $p = 0.529$).

Finally, we collected users' historical tweets—including posts, retweets, and quotes—which span two months before posting tagged tweets and two months after exposure to misinformation tagging, resulting in 1,409,845 tweets in total. Posts typically indicate active engagement with diverse political sources and topics, allowing users to express their opinions. In contrast, retweets and quotes—which involve sharing others' tweets—suggest more passive engagement, not necessarily reflecting personal views. We utilize these three types of behaviors for a more comprehensive measurement of users' information engagement[60,61]. We assume that individual misinformation taggings are exposed to users when they are posted, and collective misinformation taggings are exposed to users when they are made public following the above-threshold helpfulness votes. For our statistical analyses, we included 712,948 tweets with observed political and content diversity scores. This research study received a determination from the University of Chicago Social & Behavioral Sciences Institutional Review Board that the study is not considered human subjects research and does not require review (Institutional Review Board Protocol IRB24-0051).

### Political diversity

Political diversity measures whether a user posted a tweet that referenced sources having an opposite political stance. Specifically, we determine the political stance of the referenced source by extracting the domain (e.g., cnn.com) of the source and check it from MediaBias/FactCheck database (MBFC; https://mediabiasfactcheck.com/)[5,47]. MBFC provides a continuous score for 4874 websites to indicate each source's political stance, ranging from -1 (extreme left) to 1 (extreme right). Our additional analysis shows that political stance scores from MBFC show significant inter-rater reliability with another database of the political stance of news media, AllSides.com (see Supplementary Method 8).

We then calculate a user's political stance by averaging the political stance scores of sources referenced in their historical tweets which span two months before posting tagged tweets and two months after misinformation tagging (see Supplementary Fig. 4). Users who predominantly cite left-leaning media are considered left, and those who cite right-leaning media are considered right. Specifically, users with negative average political stance scores are categorized as left, while those with positive scores are categorized as right. Finally, we assign a binary value to represent a user's political diversity: 1 (diverse) if a user cited a source that has an opposite political stance from the user's own political stance, 0 (not-diverse) if a user cited a source with the same political stance.

The mean political diversity score is 0.166, and the standard deviation is 0.372 ($N = 712,948$). Political diversity is negatively correlated with the number of tweets posted per day ($r(712946) = −0.107$, 95% CI = [−0.109, −0.105], $p < 0.001$) and the proportion of retweets ($r(712946) = −0.046$, 95% CI = [− 0.048, −0.044], $p < 0.001$). This indicates that users are more active within information bubbles, actively posting tweets rather than passively retweeting other users' tweets within these bubbles (see Supplementary Table 2).

### Content diversity

Content diversity measures whether a user posted a tweet with a topic that is rarely discussed in the user's historical tweets. We apply the Twitter4SSE model, a transformer-based sentence embedding model (SentenceBERT) that was initialized from BERTweet (a RoBERTa model trained on 850 million tweets from 2012/1 to 2019/8 and 5 million tweets related to COVID-19 pandemic), to encode the meaning of a tweet into a 768-dimensional vector[62,63]. The model was further optimized based on recent data (75 million tweets from 2020/11 to 2020/12) using Multiple Negatives Ranking Loss (MNRL) to identify semantic similarity based on the principle that tweets quoting or replying to the same original tweet are likely discussing related ideas[62]. If a pair of tweets quoted or replied to the same tweet, the semantic similarity between them is assumed to be high.

To apply the Twitter4SSE model, we first conduct the identical data preprocessing steps to clean the tweets, which includes: eliminate URLs and mentions and transform the text to lowercase to reduce the presence of generic texts[62]. Next, we represent each tweet with a 768-dimensional semantic embedding (Supplementary Fig. 5 shows the visualization). Finally, we measure the cosine distance between the user embedding and tweet embedding (see Fig. 1b) to represent the content diversity of the current tweet. The user embedding is the average embedding of the user's historical tweets (see Fig. 1b). Estimating the distance in the embedding space has been frequently used to quantify the diversity of user activities in the online platform[48,64]. The distance ranges from 0 to 0.835, with 0 representing homogeneous content and 0.835 representing extremely diverse content. The mean content diversity score is 0.357, and the standard deviation is 0.109 ($N = 712,948$). We find that political and content diversity are slightly correlated ($r(712946) = 0.020$, 95% CI = [0.018, 0.022], $p < 0.001$), assessing conceptually distinct aspects of diversity.

Table 1 shows an example of how content diversity scores are assigned. In this example, the user primarily shows interests in COVID-19 related misinformation. However, as the user explores diverse topics—tax, LGBTQ+, international issues, and labor—the content diversity score increases.

Content diversity is negatively correlated with the number of tweets posted per day ($r(712946) = −0.052$, 95% CI = [− 0.054, −0.050], $p < .001$) but positively correlated with the proportion of retweets ($r(712946) = 0.012$, 95% CI = [0.010, 0.014], $p < 0.001$). In other words, users tend to retweet others' tweets rather than posting their own tweets when increasing content diversity (see Supplementary Table 2).

### Interrupted time series (ITS) analysis

We apply Interrupted Time Series (ITS) analysis to examine how individual and collective misinformation tagging affect the trend of political and content diversity in posting behavior. We fit the ITS model to the time series around fact-checking events, spanning five weeks (35 days) before posting the fact-checked tweet and five weeks after fact-checking. To compare the differential impacts of individual and collective misinformation tagging, we formulate the following multi-group ITS model. We control for user fixed effects to eliminate the user-related unobserved time-invariant heterogeneity that could possibly affect the outcomes. Additionally, the number of tweets posted per day is negatively correlated with political diversity ($r(712946) = −0.107$, 95% CI = [−0.109, −0.105], $p < 0.001$) and content diversity ($r(712946) = −0.052$, 95% CI = [− 0.054, −0.050], $p < 0.001$), indicating that users are more active within information bubbles. Therefore, we control for the number of tweets posted per day to ensure that our analysis focuses on variations in diversity rather than engagement volume.

For each tweet, let $Y$ be the outcome variable (i.e., political or content diversity of a specific tweet), $W$ is the weeks before posting the tweet with misinformation (negative values) or after misinformation tagging (positive values). Note that we measure $W$ by dividing the days by 7. For example, if a particular tweet is posted 3 days before posting the tweet, $W$ is $-3/7$. $T$ is an indicator of the treatment status where 0 represents a tweet posted before misinformation tagging and 1 represents after tagging. $C$ is an indicator of the type of misinformation tagging where 0 represents individual tagging and 1 represents collective tagging. $N$ corresponds to the number of tweets per day (control variable), $\alpha$ corresponds to the user fixed effect, and and $\epsilon$ is the error term. Then the ITS model is defined:

$$Y = \beta_0 + \beta_1 W + \beta_2 T + \beta_3 WT + \beta_4 WC + \beta_5 TC + \beta_6 WTC + \beta_7 N + \alpha + \epsilon \quad (1)$$

Here, $\beta_0$ is the intercept, $\beta_1$ is the slope before individual misinformation tagging. $\beta_2$ is the change in the outcome immediately after the individual misinformation tagging. $\beta_3$ is the slope change before and after individual misinformation tagging. $\beta_1 + \beta_4$ is the slope before collective misinformation tagging. $\beta_2 + \beta_5$ is the change in the outcome immediately after the collective misinformation tagging. $\beta_3 + \beta_6$ is the slope change before and after collective misinformation tagging. Thus, $\beta_4$, $\beta_5$, $\beta_6$ are the terms that estimate the differences between the effects of individual and collective misinformation tagging. Supplementary Table 12 shows how these estimates correspond to each cell in Table 1 for each outcome.

Before estimating the model, political diversity (binary variable) has been multiplied by 100 so that the coefficients are interpretable as absolute percentage point changes. Content diversity has been normalized to z-scores (i.e., the number of standard deviations from the mean). When estimating the statistical significance of the estimates, all $p$-values are two-sided. The thresholds for statistical significance is set at $p < 0.05$, and marginal significance is set at $p < 0.1$.

### Delayed feedback (DF) analysis

In addition to the interrupted time series (ITS) analysis, we conduct a delayed feedback (DF) analysis to estimate the causal impacts. We begin by establishing control and treatment groups: each tweet is paired with another tweet that was subject to misinformation tagging at an earlier time. Specifically, for every tweet in a control group, we search for a corresponding treatment tweet using the following criteria: (1) They must have been fact-checked using the same approach, either individual or collective misinformation tagging. (2) They should have been fact-checked prior to the control tweet. (3) They should have the same topic, considering that distinct topics of misinformation could lead to different levels of political and content diversity (see "Method": Topic Modeling for a detailed explanation of the topic modeling process). (4) They should have been posted no more than seven days apart from the control tweet. In cases where we have multiple tweets that meet these criteria, we choose the one with the closest posting time to the control tweet. This results in 476 pairs of tweets in control and treatment groups.

For each pair of tweets, we identify two time windows: pre-treatment ($t_0$) and post-treatment ($t_1$). $t_1$ represents the time window when the treatment tweets are fact-checked but the control tweets are not. If the duration of $t_1$ exceeds a seven-day window, we use the data within the seven-day window after receiving the tags, considering that the timing of the fact-check could affect the outcome. $t_0$ represents the time window (with equal duration of $t_1$) when both the treatment and control tweets are not fact-checked. Then we design the following difference-in-differences model to assess the impacts of misinformation tagging.

For each tweet, let $Y$ be the outcome variable (i.e., political or content diversity of a specific tweet). $T$ is a binary variable indicating whether the treatment tweet, but not control tweet, receives the treatment (i.e., misinformation tagging). $G$ is a binary variable indicating whether the tweet is assigned in the treatment (i.e., 1) or control group (i.e., 0). $C$ is an indicator of the type of misinformation tagging where 0 represents individual tagging and 1 represents collective tagging. $N$ corresponds to the number of tweets per day (control variable), $\alpha$ corresponds to the user fixed effect, and and $\epsilon$ is the error term.

$$Y = \beta_0 + \beta_1 TG + \beta_2 TGC + \beta_3 T + \beta_4 TC + \beta_5 N + \alpha + \epsilon \qquad (2)$$

$\beta_0$ is the intercept. $\beta_1$ is the difference in pre-post change in the outcome between the control and treatment group for individual misinformation tagging. $\beta_1 + \beta_2$ is the difference in pre-post change for collective misinformation tagging. Thus, $\beta_2$ is the term that estimates the difference between the effects of individual and collective misinformation tagging. $\beta_3$ and $\beta_4$ account for the baseline changes in the outcomes. Supplementary Table 13 shows how these estimates correspond to each cell in Table 2 for each outcome.

Like ITS models, political diversity (binary variable) has been multiplied by 100 so that the coefficients are interpretable as absolute percentage point changes. Content diversity has been normalized to z-scores (i.e., the number of standard deviations from the mean). When estimating the statistical significance of these coefficients, all p-values are two-sided. The thresholds for statistical significance is set at $p < .05$, and marginal significance is set at $p < 0.1$.

Supplementary Fig. 6 illustrates day-to-day changes in the outcome variables (political and content diversity) for the control and treatment groups over the pre-intervention period, controlling for time-invariant differences across users (i.e., user fixed effects). Supplementary Fig. 6 suggests that control and treatment groups follow similar trends in the absence of the tagging intervention. To statistically test this, we fitt a linear model: $Diversity = \beta_0 + \beta_1 Day + \beta_2 IsTreatmentGroup + \beta_3 Day \cdot IsTreatmentGroup + \varepsilon$, where $Day$ is the number of days in the matched pre-intervention period, and $IsTreatmentGroup$ is 1 for the treatment group and 0 for the control group. If the parallel trends assumption holds, we would find a non-significant slope difference ($\beta_3$). Our results show no significant slope difference for both political diversity ($\beta_3 = -0.493$, 95% CI = [−1.644, 0.725], $t(3) = -0.864$, $p = .408$) and content diversity ($\beta_3 = 0.011$, 95% CI = [−0.036, 0.058], $t(3) = 0.511$, $p = 0.620$) (see Supplementary Table 14).

Supplementary Fig. 7 illustrates average baseline changes of political and content diversity obtained from the control group in the DF analysis. According to the baseline changes, we find that political diversity significantly increases ($\beta = 1.956$, 95% CI = [0.068, 3.843], $t(8182) = 2.03$, $p = 0.042$), but content diversity does not significantly decrease ($\beta = -0.039$, 95% CI = [−0.079, 0.002], $t(8182) = -1.88$, $p = 0.060$) if the problematic tweet is not tagged. In other words, we find that political diversity consistently increases over time without tagging in the control group. On the other hand, we find that content diversity does not significantly change. Our results show that the effects of individual and collective tagging are above and beyond these baseline changes (see Results: Delayed Feedback (DF) Analysis).

### Topic modeling
We apply BERTopic to extract latent topics from tweets that received misinformation tags[65]. Specifically, we first represent each tweet with a 768-dimensional semantic embedding using Twitter4SSE. Then, we map the embeddings to a 5-dimensional space via UMAP (Uniform Manifold Approximation and Projection) to mitigate the curse of dimensionality[66,67]. Next, we apply HDBSCAN (Hierarchical Density-Based Spatial Clustering of Applications with Noise) to identify clusters of topics[68]. Unlike k-means algorithms, HDBSCAN does not require the user to pre-specify the number of clusters, and HDBSCAN is adept at identifying and handling noise, distinguishing between topics and outliers, which is crucial for maintaining the integrity of the clustered topics.

Traditional methods such as LDA extract topics based on bag-of-words and often fall short when applied to short texts like tweets[69]. BERTopic emerges as particularly advantageous for analyzing data from Twitter and it preserves the semantic structure of the text[63], thus enhancing its ability for short-text analysis compared to traditional models.

We generate 23 topics for 6660 fact-checked tweets, and 1073 tweets are not assigned any topic and thus considered as outliers. These outliers are excluded from the process of assigning tweets into control and treatment groups in the delayed feedback (DF) analysis. Most frequent topics with the keywords are shown in the Supplementary Table 7. As shown in Supplementary Table 7, the nine most frequent topics in our dataset include political topics that are known to trigger divisive, polarized reactions in US politics, such as COVID-19 vaccine-related misinformation (Topic 1), election- and politician-related misinformation (Topic 4, 5, 7, 8), policy-related misinformation (Topic 3, 6), and environment and disaster-related misinformation (Topic 9). These topics account for 84.06% of the corrections made through individual tagging and 59.49% of the corrections made through collective tagging. Given the time period of collection, the most frequent topic is about COVID-19 pandemic and vaccination.

### Linguistic characteristics of misinformation tagging messages
For each misinformation tagging, we analyze the message's toxicity, sentiment, length, reading ease, and delayed response time to provide insights into the qualitative differences between individual and collective misinformation tagging. Supplementary Table 4 shows the descriptive statistics of the following variables.

- Toxicity: We apply Google Jigsaw Perspective API to measure the probability that a particular message is toxic (range from 0 to 1[10,70]).
- Sentiment: We conduct Vader sentiment analysis to estimate sentiment scores of messages (on a [−1, 1] scale[14,71]). The scale spans from -1, denoting negative sentiment, to 1, denoting positive sentiment.
- Length: We measure the length of messages based on the number of characters[14].
- Reading ease score: We evaluate the readability of messages with the Flesch-Kincaid Reading Ease score (on a [1,100] scale, where large value indicates easier readability[14,72]). The Flesch−Kincaid reading ease score was transformed into an 8-level categorical variable: "5th grade" for scores 100−90, "6th grade" for 90−80, "7th grade" for 80−70, "8th & 9th grade" for 70−60, "10th to 12th grade" for 60−50, "College" for 50−30, "College graduate" for 30−10, and "Professional" for 10−0.
- Delayed response time: We calculate it as the number of days between original tweets and misinformation tagging.

### Robustness checks
We verify our findings with a battery of robustness checks. First, there might be concerns that our conclusions about the effects of misinformation tagging on human users may be biased by the presence of bots on Twitter. Many studies have utilized bot detection algorithms to exclude users who are likely to be bots to address this concern[2,5], but others argue these algorithms lead to false negatives (i.e., bots

Article

misclassified as humans) and positives (i.e., humans misclassified as bots) that could further bias analyses, even when used cautiously[54]. To mitigate concerns of bot prevalence, we reanalyze our data excluding accounts identified as bots using BotometerLite API. Specifically, using BotometerLite API, we evaluate the likelihood of users in our dataset being bot accounts and remove 360 accounts that have a likelihood higher than 50%[73], which do not meaningfully change our results. In interrupted time series (ITS) models, after individual tagging, both political diversity ($\beta = -0.882$, 95% CI = [−1.326, −0.438], $t(394654) = -3.89$, $p < 0.001$) and content diversity ($\beta = -0.026$, 95% CI = [−0.037, −0.014], $t(394654) = -4.27$, $p < 0.001$) significantly decrease. After collective tagging, content diversity significantly increases ($\beta = 0.052$, 95% CI = [0.020, 0.084], $t(394654) = 3.16$, $p = 0.002$) (refer to Supplementary Table 15 and 10). In delayed feedback (DF) models, the effects of individual tagging on political diversity ($\beta = -5.348$, 95% CI = [−9.154, −1.542], $t(7766) = -2.75$, $p = .006$) and collective tagging on content diversity are significant ($\beta = 0.292$, 95% CI = [0.149, 0.435], $t(7766) = 4.00$, $p < 0.001$) (refer to Supplementary Table 17).

In terms of applying BotometerLite API, some features are missing in our dataset collected with Twitter API 2.0: (1) *default_profile* (whether the user altered the theme or background of their user profile); (2) *profile_use_background_image* (whether the user has a background image or not); and (3) *favourites_count* (number of likes posted by the user, which were only available in Twitter API 1.1). To address this issue, we conduct missing data imputation with the IterativeImputer in Sklearn. We train an imputation model with 90,000 tweets randomly selected in August 2021 from the Twitter Stream Grab (https://archive.org/details/twitterstream). We then evaluate the model with a held-out sample of 10,000 tweets. The model performance for predicting the missing features is as follows: default_profile at 0.95 *F-1* score, profile_use_background_image at 0.90 *F-1* score, favourites_count was 0.10 $R^2$ value. Finally, we apply this imputation model to recover the missing features in our dataset.

Second, we control for potentially insincere informational activities, such as citing sources of low credibility and intentionally spreading fake news. Some might question whether the increase in political and content diversity is associated with these insincere activities. Put simply, users might be engaging with diverse information that includes misleading claims and conspiracy theories. For each tweet posted by each poster, we measure the credibility of the referenced source. Specifically, we use the binary credibility scores (1 = low credibility; 0 = medium or high credibility) from the MediaBias/FactCheck database. Our analysis indicates a strong negative correlation between the engagement of low-credibility sources and measures of political diversity ($r = -0.227$, 95% CI = [−0.229, −0.225], $p < 0.001$) and content diversity ($r = -0.030$, 95% CI = [−0.032, −0.028], $p < 0.001$). This implies that the increase of diversity in information engagement reflects engagement with a healthier information ecosystem, rather than the reverse. Furthermore, we reassess our data while controlling for credibility of sources, and find that our results remain unaffected. In ITS models, after individual tagging, both political diversity ($\beta = -0.972$, 95% CI = [−1.395, −0.549], $t(418114) = -4.50$, $p < 0.001$) and content diversity ($\beta = -0.030$, 95% CI = [−0.042, −0.018], $t(418114) = -5.08$, $p < 0.001$) significantly decrease. After collective tagging, content diversity significantly increases ($\beta = .041$, 95% CI = [0.012, 0.070], $t(418114) = 2.76$, $p = 0.006$) (refer to Supplementary Table 15 and 10). In DF models, the effects of individual tagging on political diversity ($\beta = -4.958$, 95% CI = [−8.591, −1.324], $t(8181) = -2.67$, $p = 0.007$) and collective tagging on content diversity are significant ($\beta = 0.279$, 95% CI = [0.150, 0.407], $t(8181) = 4.25$, $p < 0.001$) (refer to Supplementary Table 17). This implies that posters' exploration prior to being tagged was likely well-intentioned and would have been efficacious had they not been prompted to retreat.

Third, we attempt to avoid situations in which posters simply criticize distant information without honest consideration by filtering out tweets with negative sentiment. For each tweet posted by each poster, we conduct Vader sentiment analysis to estimate sentiment scores (on a [−1, 1] scale[14,71]). Then we exclude tweets that have a sentimental score lower than 0, which do not meaningfully change our results, except for making the immediate change of content diversity after collective misinformation tagging not significant in ITS models ($\beta = 0.009$, 95% CI = [−0.031, 0.048], $t(262015) = 0.43$, $p = 0.666$). In ITS models, after individual tagging, both political diversity ($\beta = -1.470$, 95% CI = [−2.032, −.909], $t(262015) = -5.13$, $p < .001$) and content diversity ($\beta = -0.027$, 95% CI = [−0.042, −0.011], $t(262015) = -3.35$, $p = 0.001$) significantly decrease (refer to Supplementary Table 15 and 10). In DF models, the effects of individual tagging on political diversity ($\beta = -8.075$, 95% CI = [−12.757, −3.394], $t(4973) = -3.38$, $p = 0.001$) and collective tagging on content diversity are significant ($\beta = 0.294$, 95% CI = [0.114, 0.474], $t(4973) = 3.20$, $p = 0.001$) (refer to Supplementary Table 17).

Fourth, we address concerns regarding the possibility of miscoding mentions of "community note." Specifically, we identify all tweets that mention keywords about receiving community notes broadly (i.e., community note, birdwatch, fact-check, factcheck, politifact) within the sample and remove those tweets. To address the concern regarding the effect of replying back to individual taggers, we identify all tweets that reply directly to the individual taggers and remove them. We find that the effects on political and content diversity do not meaningfully change in both ITS and DF analyses. In ITS models, after individual tagging, both political diversity ($\beta = -1.048$, 95% CI = [−1.488, −0.608], $t(416183) = -4.67$, $p < 0.001$) and content diversity ($\beta = -0.030$, 95% CI = [−0.041, −0.018], $t(416183) = -5.00$, $p < 0.001$) significantly decrease. After collective tagging, content diversity significantly increases ($\beta = 0.041$, 95% CI = [0.012, .069], $t(416183) = 2.75$, $p = 0.006$) (refer to Supplementary Table 18). In DF models, the effects of individual tagging on political diversity ($\beta = -5.329$, 95% CI = [−9.139, −1.519], $t(8088) = -2.74$, $p = .006$) and collective tagging on content diversity are significant ($\beta = 0.276$, 95% CI = [0.148, 0.405] $t(8088) = 4.22$, $p < 0.001$) (refer to Supplementary Table 19).

Fifth, to strictly identify individual tags (i.e., PolitiFact links) that correct the original posters, we submit original posts, replies containing PolitiFact links, and the cited PolitiFact fact-checking articles to ChatGPT (gpt-4o-2024-05-13). We prompt the model to annotate whether the PolitiFact link was used to correct the original poster rather than support them. Consequently, we identify 5592 PolitiFact links out of 6760 links (82.72%) as corrective (see Supplementary Method 9). Subsequently, we limit the sample to the 5592 links identified by ChatGPT from the individual tagging data, which does not meaningfully alter the results. In ITS models, after individual tagging, both political diversity ($\beta = -1.086$, 95% CI = [−1.558, −0.614], $t(363273) = -4.51$, $p < 0.001$) and content diversity ($\beta = -0.039$, 95% CI = [−0.051, −0.026], $t(363273) = -6.11$, $p < 0.001$) significantly decrease. After collective tagging, content diversity significantly increases ($\beta = 0.041$, 95% CI = [0.012, 0.069], $t(416183) = 2.75$, $p = .006$) (refer to Supplementary Table 20). In DF models, the effects of individual tagging on political diversity ($\beta = -7.135$, 95% CI = [−11.068, −3.203], $t(7508) = -3.56$, $p < 0.001$) and collective tagging on content diversity are significant ($\beta = 0.272$, 95% CI = [0.143, .401] $t(7508) = 4.13$, $p < 0.001$) (refer to Supplementary Table 21).

Sixth, we restrict the sample to original posters who have replied to (and thereby read) the individual tags (i.e., fact-checking replies) and remove non-responders. Specifically, out of 6760 original posters who received individual tags, we remove 4288 posters who did not reply to the tags, resulting in 2472 posters. After that, we compare these 2472 posters with 973 posters who received collective tags. Even after removing the non-responders, we find that results regarding tagging's effects remain consistent. Specifically, as with the complete sample, we identically find that individual tagging causes immediate decrease in political and content diversity. After individual tagging, both political diversity ($\beta = -1.481$, 95% CI = [$-2.337$, $-0.626$] $t(164724) = -3.39$, $p = 0.001$) and content diversity ($\beta = -0.026$, 95% CI = [$-0.048$, $-0.004$], $t(164724) = -2.30$, $p = 0.021$) significantly decrease. After collective tagging, content diversity significantly increases ($\beta = .040$, 95% CI = [.010, .071], $t(164724) = 2.59$, $p = 0.010$) (refer to Supplementary Table 22)

### Reporting summary
Further information on research design is available in the Nature Portfolio Reporting Summary linked to this article.

### Data availability
The Twitter data collected and analyzed in this study have been deposited in the Open Science Foundation (OSF) database at https://doi.org/10.17605/OSF.IO/TXGSR. The data required to replicate our analyses are available in the repository. However, in accordance with Twitter's privacy policy, we cannot disclose individual-level user information or the contents of tweets. Instead, processed and anonymized data are available in the repository.

### Code availability
The code required to replicate our results is available at https://doi.org/10.17605/OSF.IO/TXGSR.

*Processing* 3982–3992 (Association for Computational Linguistics, 2019).

## Acknowledgements
We are grateful for the Discovery Partners Institute of Illinois for a grant on the topic of Misinformation Tagging to Kevin Leicht, PI. We thank Donghyun Kang, Yoosik Youm, and Byungkyu Lee for their helpful feedback on our manuscript. We also thank the members and alumni of the Knowledge Lab (University of Chicago) and Yonsei Social Networks & Neuroscience Lab for their comments.

## Author contributions
J.K., Z.W., and J.E. collaboratively conceived and designed the study, drafted, revised, and edited the manuscript. J.K. and H.S. performed the analysis. J.K., H.S., and H.L. gathered and cleaned the data. J.K. and Z.W. produced the visualizations.

## Competing interests
J.E. has a commercial affiliation with Google, but Google had no role in the design and analysis of this study. The authors declare no other competing interests.

## Additional information
**Supplementary information** The online version contains supplementary material available at
https://doi.org/10.1038/s41467-025-55868-0.

**Correspondence** and requests for materials should be addressed to James Evans.

**Peer review information** *Nature Communications* thanks Sarah Shugar, Jevin West and the other, anonymous, reviewer(s) for their contribution to the peer review of this work. A peer review file is available.

**Reprints and permissions information** is available at
http://www.nature.com/reprints

**Publisher's note** Springer Nature remains neutral with regard to jurisdictional claims in published maps and institutional affiliations.